\documentclass[]{aa}
\usepackage{graphics}
\usepackage{astron}
\newcommand{\beq}{\begin{equation}}
\newcommand{\beqa}{\begin{eqnarray}}
\newcommand{\eeq}{\end{equation}}
\newcommand{\eeqa}{\end{eqnarray}}
\begin{document}
   \thesaurus{06         
              (03.13.2;  
	       03.20.4;  
               10.08.1;  
               11.13.1;  
               12.04.1;  
               12.07.1)} 
\title{AGAPEROS: Searches for microlensing in the LMC with the Pixel
Method} \subtitle{II. Selection of possible microlensing events}

\author{A.-L. Melchior\inst{1,2}, C. Afonso\inst{3},
R. Ansari\inst{4}, {\'E.}  Aubourg\inst{3}, P. Baillon\inst{5},
P. Bareyre\inst{3}, F. Bauer\inst{3}, J.-Ph. Beaulieu\inst{6},
A. Bouquet\inst{2}, S.  Brehin\inst{3}, F.  Cavalier\inst{4},
S. Char\inst{7}, F. Couchot\inst{4}, C. Coutures\inst{3},
R. Ferlet\inst{6}, J.  Fernandez\inst{7}, C. Gaucherel\inst{3},
Y. Giraud-H\'eraud\inst{2}, J.-F. Glicenstein\inst{3},
B. Goldman\inst{3}, P. Gondolo\inst{2,8}, M. Gros\inst{3},
J. Guibert\inst{9}, D. Hardin\inst{3}, J. Kaplan\inst{2}, J. de
Kat\inst{3}, M. Lachi\`eze-Rey\inst{3}, B. Laurent\inst{3},
{\'E}. Lesquoy\inst{3}, Ch. Magneville\inst{3}, B. Mansoux\inst{4},
J.-B. Marquette\inst{6}, E. Maurice\inst{10}, A. Milsztajn\inst{3},
M. Moniez\inst{4}, O. Moreau\inst{9}, L. Moscoso\inst{3}, N.
Palanque-Delabrouille\inst{3}, O. Perdereau\inst{4},
L. Pr\'ev\^ot\inst{10}, C. Renault\inst{3}, F. Queinnec\inst{3},
J. Rich\inst{3}, M. Spiro\inst{3}, A. Vidal-Madjar\inst{6},
L. Vigroux\inst{3}, S. Zylberajch\inst{3}}

\offprints{A.L.Melchior@qmw.ac.uk}

\institute{Astronomy Unit, Queen Mary and Westfield College, Mile End Road,
London E1\,4NS, UK
\and Laboratoire de Physique Corpusculaire et Cosmologie (UMR 7535),
Coll\`ege de France, 75\,231 Paris Cedex 05, France
\and CEA, DSM, DAPNIA, Centre d'\'Etudes de Saclay, 91\,191
Gif-sur-Yvette Cedex, France
\and Laboratoire de l'Acc\'el\'erateur Lin\'eaire, IN2P3 CNRS,
Universit\'e Paris-Sud, B.P. 34, 91\,898 Orsay Cedex, France
\and CERN, 1211 Gen\`eve 23, Switzerland
\and Institut d'Astrophysique de Paris, CNRS, 98 bis Boulevard Arago,
75\,014 Paris, France
\and Universidad de la Serena, Faculdad de Ciencias, Departemento de
Fisica, Casilla 554, La Serena, Chile.
\and Max-Planck-Institut f\"ur Physik, F\"ohringer Ring 6, 80\,805
M\"unchen, Germany
\and Centre d'Analyse des Images de l'INSU, Observatoire de Paris, 61
avenue de l'Observatoire, 75014 Paris, France
\and Observatoire de Marseille, 2 place Le Verrier, 13\,248 Marseille
Cedex 04, France}

\date{Received / Accepted}
\titlerunning{Searches for microlensing in the LMC... II}
\authorrunning{A.-L. Melchior et al.}
\maketitle

\begin{abstract} 
We apply the pixel method of analysis (sometimes called ``pixel
lensing'') to a small subset of the EROS-1\ microlensing observations
of the bar of the Large Magellanic Cloud (\object{LMC}).  The pixel
method is designed to find microlensing events of unresolved source
stars and had heretofore been applied only to M31 where essentially
all sources are unresolved. With our analysis optimised for the
detection of long-duration microlensing events due to 0.01-1 $M_\odot$
Machos, we detect no microlensing events and compute the corresponding
detection efficiencies.  We show that the pixel method, applied to
crowded fields, should detect 10 to 20 times more microlensing events
for $M>0.05 M_\odot$ Machos compared to a classical analysis of the
same data which latter monitors only resolved stars.  In particular,
we show that for a full halo of Machos in the mass range 0.1 -- 0.5
$M_\odot$, a pixel analysis of the three-year EROS-1 data set covering
$0.39\,\rm deg^2$ would yield $\simeq 4$ events.

\keywords{Methods: data analysis -- Techniques: photometric -- Galaxy:
halo -- Galaxies: Magellanic Clouds -- Cosmology: dark matter --
Cosmology: gravitational lensing}
\end{abstract}

\section{Introduction}
Microlensing searches can probe the distribution of MAssive Compact
Halo Objects (Machos) in the dark halo of our Galaxy or more distant
galaxies (Paczy\'nski (1986,1996), Griest
(1991))\nocite{Paczynski:1986,Griest:1991,Paczynski:1996}. Several
lines of sight are now under investigation, and events have been
claimed in several directions: towards the \object{LMC} (Alcock et
al.\ 1997a, Renault et al.\ 1997)\nocite{Alcock:1997a,Renault:1997},
the SMC (Alcock et al.\ 1997b, Pa\-lan\-que-Delabrouille et~al.
1998\nocite{Alcock:1997b,Palanque-Delabrouille:1998}), in the
direction of the Galactic Bulge (Alard et al.\ 1997, Alcock et
al. 1997c, Udalski et al
1994)\nocite{Alcock:1997d,Alard:1997,Udalski:1994b} and more recently
towards spiral arms (Derue et al.\ 1998)\nocite{Derue:1998}, giving
some first evidences of the Macho distribution towards these lines of
sight. These results are based on a star monitoring analysis: the
fluxes of several millions of {\it resolved} stars are monitored. As
first discussed by Crotts (1992)\nocite{Crotts:1992} and Baillon et
al.\ (1993), events due to {\em unresolved} stars essentially escape
detection of these analyses. Such stars, beyond the crowding limit or
too dim to resolve, could significantly contribute to the number of
detectable events.  This is illustrated by the detection of two
variable objects in the MACHO analysis (Alcock et
al. 1997a)\nocite{Alcock:1997a} of \object{LMC} data, which could not
be resolved at their minimum luminosity. The detection of the
variation was nevertheless possible because the reference images used
to establish the catalogue of monitored stars were taken during their
maximum luminosity, when the stars were resolved. However, such events
can neither be kept for further considerations nor be included in the
computation of detection efficiencies, in the star monitoring
approach.

In this paper, we apply a pixel analysis to the EROS~91-92 data (10\%
of the whole EROS-1 CCD data set) of the \object{LMC Bar}. The idea is
to monitor the flux of all the pixels present on the images, thus
achieving a good sensitivity to the whole stellar content of the
images. The magnification of one unresolved star can be detected as a
variation of the pixel flux, provided that the magnification is high
enough. -- In the following, we will refer to this approach as pixel
monitoring, as opposed to star monitoring referring to classical
analyses restricted to resolved stars. -- The main uncertainties of
this approach concern our ability to account properly for variations
of the observational conditions, and to be able to disentangle
intrinsic variations from observational systematics.  In Paper~I
(Melchior et al. 1998a)\nocite{Melchior:1998a}, dedicated to the
description of the treatment of the data and the production of $2.1
\times 10^6$ super-pixel light curves, we have shown that an average
stability of $1.8\%$ of the super-pixel flux is achieved in blue and
$1.3\%$ in red, about twice the expected photon noise.  This
homogeneous set of super-pixel light curves is called AGAPEROS: each
of these light curves covers a period of $120$ days and is composed of
about 90 measurements. With this rather short period of observation,
we show here how it is possible to investigate the Macho mass range of
interest ($M\simeq 10^{-2} - 1 M_{\odot}$) with the existing EROS-1
CCD data set, initially designed for short time scale events in a mass
range ($M\simeq 10^{-8} - 10^{-3} M_{\odot}$) where no event has been
detected.

\subsection*{Microlensing selection with the Pixel Method}
\label{section:pixmeth}
 We first present the simple formalism used to describe the pixel
events in which we are interested.  The pixel flux $\phi^{p}$,
affected by a microlensing event, can be written as:
\begin{equation}
\phi^{p}_i = \alpha_i A(t_i) \phi^*_i + \phi^{\rm bg}_i
\label{eq:phi}
\end{equation}
where $i$ is the measurement number, $t$ is the time, $\alpha$ is the
seeing fraction, $A$ the magnification, $\phi^*$ the flux of the star
of interest at rest (i.e., unmagnified), and $\phi^{\rm bg}$ includes
the sky and stellar backgrounds. The magnification (Paczy\'nski
1986)\nocite{Paczynski:1986} depends on the normalized impact
parameter $u(t_i)$:
\begin{equation}
A(t_i) = \frac{u^2+2}{u\sqrt{u^2+4}}
\end{equation}
with 
\begin{eqnarray}
u(t_i) = \sqrt{u_0^2 + {\left(\frac{v_T}{R_E}\right)}^2 (t_i - t_0)^2}
\nonumber
\end{eqnarray}
where $v_T$ is the Macho transverse velocity, $u_0$ and $t_0$ the
impact parameter and time of closest approach, and $R_E$ the Einstein
radius:
\begin{eqnarray}
R_E = \sqrt{\frac{4 G M}{c^2} \frac{D_{OL} D_{LS}}{D_{OS}}}\nonumber
\end{eqnarray}
The typical time scale of the variation is the Einstein radius
crossing time:
\begin{equation}
t_E = \frac{R_E}{v_T}
\end{equation}

In Paper~I, we showed that the variations of the observational
conditions, which obviously affect Eq.~\ref{eq:phi}, can be corrected:
each image is first geometrically aligned with the reference image.
Then sky background and absorption factor are corrected to the values
of the reference image.  Finally, each super-pixel flux is adjusted to
take account of the seeing variation, affecting $\alpha_i$.  Since the
mean seeing is about $3$ arc-second, we estimate $\alpha$ to be on average
$0.7$ for the corrected $3.\hskip-2pt ''6 \times 3.\hskip-2pt ''6$
super-pixel light curves, obtained in Paper~I.  We showed that in the
absence of microlensing events ($A\left( t_i \right) = 1$), we achieve
a proper understanding of the errors affecting these light curves, and
that Eq.~\ref{eq:phi} describes to a good approximation the light
curves we are studying.

The usual requirements used for the selection of microlensing events
detected by star monitoring can be applied here:
\begin{itemize}
\item As the microlensing phenomenon is a transient and rare
phenomenon, it should produce a {\em unique} significant variation in
the star flux. 
\item It must be {\em achromatic}. This characteristic has two
applications for a pixel analysis: the time of the maximum has to
be the same in both colours and the ratio
\begin{equation}
\frac{ {{\phi^p_i}^{(B)}} - \phi^{(B)}_{\rm bg}}{{\phi^p_i}^{(R)} -
\phi^{(R)}_{\rm bg}}
\label{eq:ach}
\end{equation}
must remain constant, during the variation.
\item The {\em shape} must be compatible with Eq.~\ref{eq:phi}. 
\end{itemize}
The first criterion allows us to remove recurrent variable stars as
well as most of the noise, while the two other criteria will be
applied to the few remaining light curves at the end of the selection
process.
\begin{itemize}
\item Last, we consider the fact that the probability for a star to be
lensed is independent of its type. This will allow us to reject
specific populations of variable stars.
\end{itemize}

In Sect. \ref{section:simul}, we present the Monte-Carlo simulations
used in this article. Then, in Sect. \ref{section:trigger}, we
describe step by step the selection procedure designed to detect
microlensing events and applied to these light curves. In
Sect. \ref{section:colmag}, we show how the selected variations can be
eliminated as compatible with variable stars. In
Sect. \ref{section:detect}, we discuss the detection efficiencies
achieved by this analysis, and compare the number of expected events
with the sensitivity of star monitoring. We rely on these results in
Sect. \ref{section:perspec} to discuss the possible prospects of this
approach.

\section{A useful tool: mock super-pixel light curves with
microlensing events} 
\label{section:simul}
Monte-Carlo simulations described in Baillon et al.\
(1993)\nocite{Baillon:1993} gave a first estimate of the number of
events expected with a pixel analysis. The main uncertainties
discussed there derived from the noise present in real data.  In
Sect. \ref{sect:phys}, we present a summary of the simulations of
microlensing events used in this paper.  These model the physical
ingredients including the halo density profile, the luminosity
function of source stars, and the Macho velocity distribution and mass
function (see Baillon et al.\ 1993). In Sect. \ref{sect:mini}, we
define a minimal threshold that will be useful later on for the
interpretation of the results of our analysis. In
Sect. \ref{sect:preli}, we discuss the characteristics of the
simulated events as expected for an ideal analysis of our data
set. Last, in Sect. \ref{sect:label}, we add to this model the
characteristics of the AGAPEROS data, and thus produce realistic mock
super-pixel light curves. This tool will be used in the following to
adjust the selection criteria in Sect. \ref{section:trigger} and to
compute the detection efficiencies in Sect. \ref{section:detect}.

\subsection{Physical ingredients}
\label{sect:phys}
We assume an isothermal halo with a core radius of $5$ kpc, normalized
at the solar neighbourhood to $\rho_\odot = 8. \times 10^{-3}
{M_\odot} pc^{-3}$ (Flores 1988)\nocite{Flores:1988} and filled with
compact objects with a given mass $M$ as discussed by Griest
(1991)\nocite{Griest:1991}. We adopt a \object{LMC} distance of $50$
kpc.  The corresponding optical depth for a full halo is the same as
the one used by the MACHO group (Alcock et al.\
1997a)\nocite{Alcock:1997a}\ $\tau_{full}~=~4.7~\times 10^{-7}$.  It
is to be noted that our estimate of the expected number of events
assumes a full halo and it should be multiplied by a factor $f<0.5$
according to the MACHO and EROS results (Alcock et al.\ 1997a, Renault
et al.\ 1997)\nocite{Alcock:1997a,Renault:1997}, where $f$ is the halo
fraction actually filled with Machos.  According to the MACHO results
(Alcock et al. 1997a)\nocite{Alcock:1997a}, it is most probably
smaller than $0.5$.  Note that no halo flattening has been considered
at this stage, but more sophisticated models could be implemented and
tested once serious candidates are detected.

We calculate the number of potential lenses with a fixed mass $M$
located in the cone pointing towards our field of view.  We assign a
random position to the Macho and choose its transverse velocity $v_T$
from a two-dimensioned Maxwellian distribution $g(v_T)$ weighted by
$v_T$.  For each given Macho, we determine the probability that a star
will lie close enough to this line of sight to give rise to a
microlensing event. We use the luminosity function described in
Baillon et al.\ (1993)\nocite{Baillon:1993} which is based on Hardy et
al.\ (1984)\nocite{Hardy:1984} for the bright stars, on Ardeberg et
al.\ (1985)\nocite{Ardeberg:1985} up to magnitude $V=23$ and is
extrapolated to the faint end using the luminosity function of the
solar neighbourhood (Allen 1973)\nocite{Allen:1973}.  Note that on the
one hand, the details of this latter extrapolation are not important
because, as we show below, few events are detectable for sources
fainter than $V=24$; but on the other hand, the connection between
these 3 sets of observations is a source of uncertainties.
\begin{figure}
\resizebox{\hsize}{!}{\includegraphics{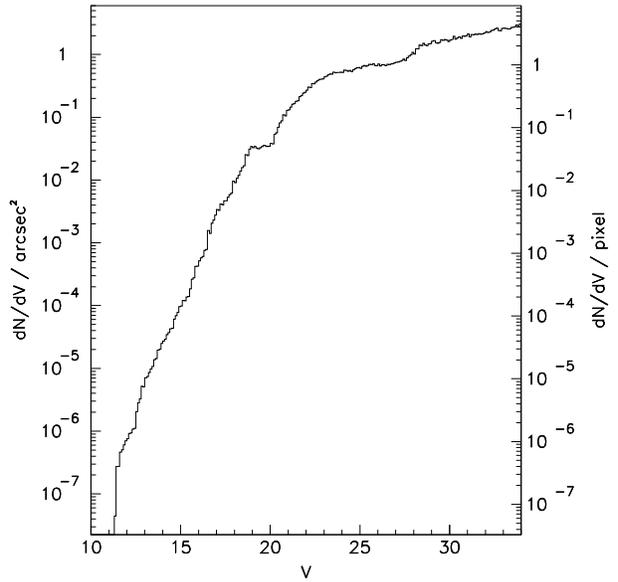}}
\caption{Luminosity function used for the simulations: the number of
stars per arcsec$^2$ or per pixel ($1.\hskip-2pt ''21 \times
1.\hskip-2pt ''21$), normalised to a surface brightness \protect
$\mu_V = 21$ is exhibited as a function of the V magnitude. This shows
the typical stellar content of a pixel.}
\label{fig:lf}
\end{figure}
This luminosity function is displayed in Fig.~\ref{fig:lf} and is
quite compatible with recent measurements (Holtzman et al.\ 1997,
Ardeberg et al.\ 1997)\nocite{Holtzman:1997,Ardeberg:1997}. We then
normalize this function to a surface brightness of $\mu_V = 21$ (de
Vaucouleurs 1957)\nocite{deVaucouleurs:1957}. The star's magnitude is
drawn from a uniform distribution and the simulated event is then
weighted according to the luminosity function.  As described in
Baillon et al.\ (1993)\nocite{Baillon:1993}, we account for possible
finite source effects that are expected when the stellar radius
projected onto the plane of the Macho is comparable with the Einstein
radius. We are then able to compute the number of expected events
using well-known Monte-Carlo integration techniques.

\subsection{Minimal requirements for simulated microlensing events}
\label{sect:mini}
We require here some minimal requirements that will define a set of
simulated microlensing events that could be detected with an ideal
experiment.

As we do not expect to detect a significant number of
low-magnification events, we introduce a threshold $A_{max}>1.34$ in
our simulations. Detectable low-magnification events affect bright
stars and hence would have already been detected by the previous EROS
star monitoring analysis anyway. Moreover, the magnification by such a
small factor of a dim star would be completely buried into the noise,
therefore one must add a visibility condition. We choose the
following: {\it at the time of maximum magnification}, the flux of the
central super-pixel of a magnified star should rise higher than
$3\sigma$ above the background, $\sigma$ being taken as twice the
photon noise. It is important to note that this threshold does not
depend on the duration of the event. It only removes events that we
would not detect in any case.

\begin{figure}
\resizebox{\hsize}{!}{\includegraphics{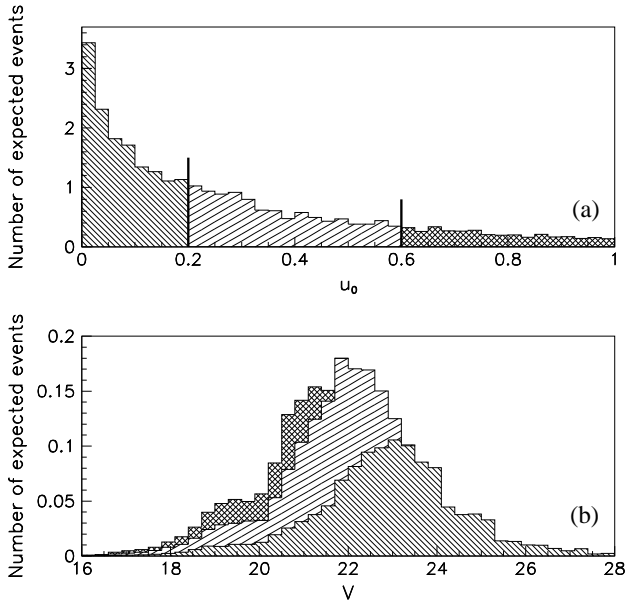}}
\caption{Initial set of the simulated light curves with some
microlensing events. The distributions of the impact parameters $u_0$
(panel {\bf a}) and of the $V$ magnitude of the un-magnified star
(panel {\bf b}) are exhibited. The histograms shown are computed for
0.5 $M_\odot$ Machos. At this stage simulations with different Macho's
mass will give the same histograms but with a different
normalization.}
\label{fig:para0}
\end{figure}

The effect of this requirement on the characteristics of the simulated
sample can be seen in Fig.~\ref{fig:para0}. The impact parameter
distribution is no longer expected flat. This $3\sigma$ threshold
introduces a (necessary) bias into the impact parameter distribution
towards small values. The majority of the events are expected to
affect dim stars with a small impact parameter $u_0 < 0.2$.

The simulations including these two thresholds ($A>1.34$, and S/N $> 3
\sigma$) will be used as a reference for the computation of our
detection efficiency in Sect. \ref{section:detect}.

\subsection{Preliminary estimates}
\label{sect:preli}
Firstly, we discuss the number of microlensing that can be expected
with an ideal experiment. Secondly, we estimate the number of stars
effectively monitored in our reference set.

\subsubsection{Number of expected events}
The number of microlensing events estimated by our simulations for a
halo filled with Machos of mass $M$ is thus defined as:
\begin{eqnarray}
N_{\rm evt} = \int d\phi^* \Biggl[ N_{\rm stars}(\phi^*)\times T_{\rm
obs} \int dv_{T} \biggl[ v_T  g(v_T) \times \nonumber \\ \int dD
\Bigl[ 2 u_0(\phi^*) R_E(D)  \frac{f\ \rho(D)}{M}  \times 
\epsilon(\phi^*,t_E(M,D,v_T)) \Bigr] \biggr] \Biggr]
\label{eq:nevt}
\end{eqnarray}
where $N_{\rm stars}(\phi^*)$ is the true number of stars with a flux
between $\phi^*$ and $\phi^* + d \phi^*$ present in the sky area
studied, $T_{\rm obs}$ is the duration of the observations (120 days),
$\rho(D)$ the Macho density distribution, $D$ the position of the
Macho and $\epsilon$ is the efficiency ($\epsilon = 1$ for an ideal
experiment).
\begin{table}
\caption[ ]{Number of expected microlensing events as a function of
the Macho mass, given the minimal requirements discussed in the text.}
\label{tab:simu2}
\begin{flushleft}
\begin{tabular}{lccccc}
\hline\noalign{\smallskip} 
$M/M_\odot$ & $0.01$ & $0.05$ & $0.1$ & $0.5$ & $1.0$
\\\noalign{\smallskip}\hline\noalign{\smallskip}   
$ N^{\rm AGAPEROS}_{\rm evt} / f$ & $ 9.8$ & $4.4$ & $3.1$ & $1.4$ &
$1.0$\\ 
\noalign{\smallskip}\hline
\end{tabular}
\end{flushleft}
\end{table}

Tab.~\ref{tab:simu2} gives the number of microlensing events that can
be expected with an ideal analysis of the super-pixel light curves
produced in Paper I. In the large-mass range, where all the known
microlensing candidates on the \object{LMC} have been identified, we
expect between 1 and 10 microlensing events assuming a full halo.

\subsubsection{Effective number of monitored stars}
If we were able to select all light curves of our reference sample, we
would define the equivalent number of monitored stars $N^{\rm
AGAPEROS}_{\rm stars}$ as follows:
\begin{eqnarray}
N_{\rm stars}^{\rm AGAPEROS} \equiv \int d\phi^* N_{\rm
stars}(\phi^*) u_0 (\phi^* \vert 3\sigma)
\label{eq:nstar}
\end{eqnarray}
$u_0 (\phi^* \vert 3\sigma)$ is the threshold impact parameter that
enters Eq.~\ref{eq:nevt} which accounts for the $3\sigma$ deviation
imposed at the time of maximum magnification in
Sect. \ref{sect:mini}. Actually, this number $N^{\rm AGAPEROS}_{\rm
stars}$ would only depend on the luminosity function and the
definition of our reference sample. Hence, we can consider that we
effectively monitor the equivalent of $2.2 \times 10^6$ stars, whose
mean magnitude is $22.1$ (see Fig.~\ref{fig:para0}b).

 If one integrates the luminosity
function\footnote{We estimate that the remaining uncertainty on the LF
translates into 20\% uncertainty in this effective number of stars.}
(Fig.~\ref{fig:lf}) over a pixel area ($1.\hskip-2pt ''21 \times
1.\hskip-2pt ''21$), one finds one star in the magnitude range $20 -
24$, that could undergo a microlensing event. This explains why the
effective number of stars thus defined is of the same order as the
number of pixels.

\subsection{Model of the data}
\label{sect:label}
The idea is to simulate super-pixel light curves that include
microlensing events. We compute the flux of the star -- affected by a
microlensing variation -- which enters the $3.\hskip-2pt ''6 \times
3.\hskip-2pt ''6$ super-pixel.  We also add a background flux --
together with expected read-out and photon noises -- in order to
obtain realistic mock light curves.  The computation of these fluxes
takes account of the pass-band of the filters, the quantum efficiency
of the CCD camera and its gain. (See Paper~I and references therein
for more quantitative information about the characteristics of the raw
data.)  Actual spacing and variations of the observational conditions
(absorption and sky background), measured on the data, are also taken
into account in this procedure. Similarly to what is done for real
data, the averaging procedure of the measurements available each night
is applied to these mock light curves, as well as to the computation
of error bars.  We multiply these errors, assumed to be gaussian and
uncorrelated, by the factor found in Paper I between the measured
dispersion and the expected photon noise.

\begin{figure}
\resizebox{\hsize}{!}{\includegraphics{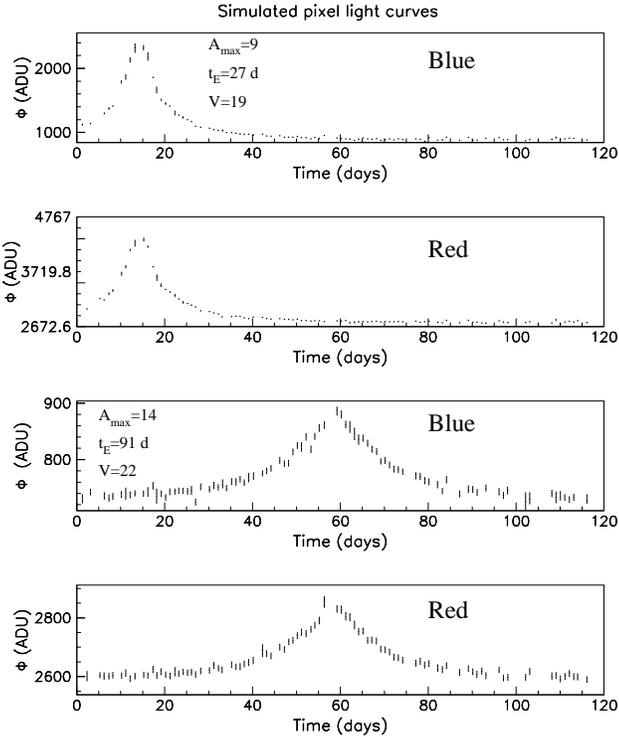}}
\caption{Two examples of simulated super-pixel light curves containing
a microlensing event.}
\label{fig:clsimu}
\end{figure}

We finally get mock light curves typical of the microlensing events we
are looking for. Fig.~\ref{fig:clsimu} displays two examples of
typical expected events.

\section{Selection of microlensing events}
\label{section:trigger}
We apply a pixel analysis designed to select microlensing events on
the EROS 91-92 data of the \object{LMC}.  Using the methods described
in Paper~I and applied on the AGAPEROS data set, we constructed some
$2.1 \times 10^6$ ({\it real}) super-pixel light curves, cleaned of
all observational effects and corrected for systematic effects to the
degree possible. In this section, we define a selection process
designed to select possible microlensing events.  The application of a
basic trigger -- detection of bumps -- reveals a large number of
variations, most of them corresponding to obvious variable stars, but
also to some noisy variations that we will have to eliminate.  Owing
to the averaging of the images taken each night (see Paper~I), some of
the variations are most probably due to short-time scale variables
already studied elsewhere (Beaulieu 1995, Grison
1995)\nocite{Beaulieu:1995b,Grison:1994b}. As Renault et al.
(1997,1998)\nocite{Renault:1997,Renault:1998} have already excluded
the small-mass Macho range with this data set, we choose to optimise
our sensitivity to long time-scale variations, corresponding to the
large-mass range in which all the known microlensing events have been
detected.

The selection criteria should remove intrinsic variations and
systematic effects, keeping genuine microlensing events.  We describe
these various criteria successively applied to the data.  Some rather
loose cuts, applied on super-pixel light curves, turn out to be
sufficient to reduce considerably the number of light curves to
analyse. Then a visual inspection of the 120 remaining curves confirms
that they are affected by genuine variations and they will be further
studied in Sect. \ref{section:colmag}. The efficiency of our selection
procedure with simulated super-pixel light curves has been checked in
both colours.

We work on a simulation based on $5\times 10^6$ realizations, which
allows to have small statistical errors on the number of expected
events.  The number of events given at each step of the selection
procedure is calculated assuming a halo full of $0.5 M_\odot$ Machos.
In Sect. \ref{section:detect}, we discuss the sensitivity actually
achieved in a larger-mass range $0.01 M_\odot \le M \le 1 M_\odot$.

The selection procedure, defined in this section and summarised in
Tab.~\ref{tab:sum}, splits into $3$ steps: (1) a significant variation
must be present in at least one colour. (2) Light curves with any
significant secondary variations are eliminated. (3) A correlation
between the two colours is required. The two first thresholds are based
on the following definition of the variations or ``bumps''.
\begin{table}
\caption[ ]{Summary of the selection procedure: the percentage of
events kept at each steps is given for the simulations (2nd column)
and the data (4rd column). The 3rd and 5th columns give the number of
simulated and real light curves kept.}
\label{tab:sum}
\begin{flushleft}
\begin{tabular}{lcccc}
\hline
\noalign{\smallskip}
 &  \multicolumn{2}{c}{Simulations}  & \multicolumn{2}{c}{Data} \\
\noalign{\smallskip}  \noalign{\smallskip}
Starting from & \multicolumn{2}{c}{$1.4$} &
\multicolumn{2}{c}{$2.1\times 10^6$} \\ 
& \multicolumn{2}{c}{events} & \multicolumn{2}{c}{light curves} \\  
\noalign{\smallskip} \hline\noalign{\smallskip}
One bump        &  \multicolumn{4}{c}{}\\
\hspace{0.1cm}- $L_1>500$      & 51.0\% & 0.71 & 0.25\%  &  
5\,338  \\
\hspace{0.1cm}- 3 points above 3$\sigma$ & 71.2\% & 0.51 & 52.3\% & 2\,789
\\\noalign{\smallskip} No second bump & 90.3\% & 0.46 & 77.5\% & 2\,162 \\
Good correlation& 85.5\% & 0.39 & 5.5\% & 120 \\ \noalign{\smallskip} \hline
\end{tabular}
\end{flushleft}
\end{table}

\paragraph*{Definition of a bump}\hspace{0.1cm}
A baseline $\phi_{bl}$ is calculated for each super-pixel light curve
as the minimum of a running average over 5 successive flux
measurements. $\sigma_{bl}$ is the error associated with the baseline
flux determination.  All the light curves are scanned for the
detection of bumps, defined as at least $3$ consecutive measurements
lying above the baseline by at least $3 \sigma_n$
\begin{equation}
\sigma_n = \sqrt{{\sigma^{\prime}_{n}}^2 + {\sigma_{bl}}^2}
\end{equation}
where ${{\sigma}^{\prime}_{n}}$ is the error associated to each
super-pixel flux computed in paper I for the night $n$. The bump ends
when at least 2 consecutive measurements lie below this threshold.
Each bump $i$ is characterised using a likelihood function:
\begin{equation}
L_\mathrm{i} = -\ln \left(\prod_{n \in \mathrm{bump~i}} P(\phi \ge
\phi_n)\ \mbox{ given }\ \left\{
\begin{array}{c} \phi_{bl} \\ \sigma_n \end{array} \right. \right)
\label{lykelyhood}
\end{equation}
$\phi_n$ is the super-pixel flux for the measurement $n$, whose
computation for all the light curves is detailed in Paper~I.

\subsection{At least one significant variation}
Firstly, we require one significant bump in at least one colour. Secondly,
we then look for a minimal variation in the other colour.

\paragraph*{A large bump in at least one colour}\hspace{0.5cm}
In order to identify significant variations, we require the likelihood
function $L_1$ associated with the largest fluctuation to be larger
than $500$ {\em for at least one colour}. This value is chosen using
the Monte-Carlo simulations to optimise the S/N ratio.  In a given
colour, we search for clusters of super-pixels\footnote{We use a
Friend of Friends algorithm (see for instance Huchra \& Geller
(1982))\nocite{Huchra:1982}.} having each $L_1$ larger than 500. We
then select the {\em central} super-pixel of each cluster, if it also
satisfies the $L_1> 500$ requirement. The latter requirement is
intended to remove some artifacts, in particular close to bright
stars.  The next cuts will be applied on these central super-pixels.

\paragraph*{Minimal variation in the other colour}\hspace{0.5cm}
The previous cut has allowed to detect significant variations in at
least one colour. We now require that the first fluctuation of the
central super-pixels of the other colour to have at least $3$
consecutive points $3 \sigma$ above the baseline. Although it
corresponds to a quite small value of $L$ ($\simeq 15$), it
constitutes a first requirement of achromaticity.

\subsection{No significant second bump}
\label{sect:2ndcol}
\begin{figure}
\resizebox{\hsize}{!}{\includegraphics{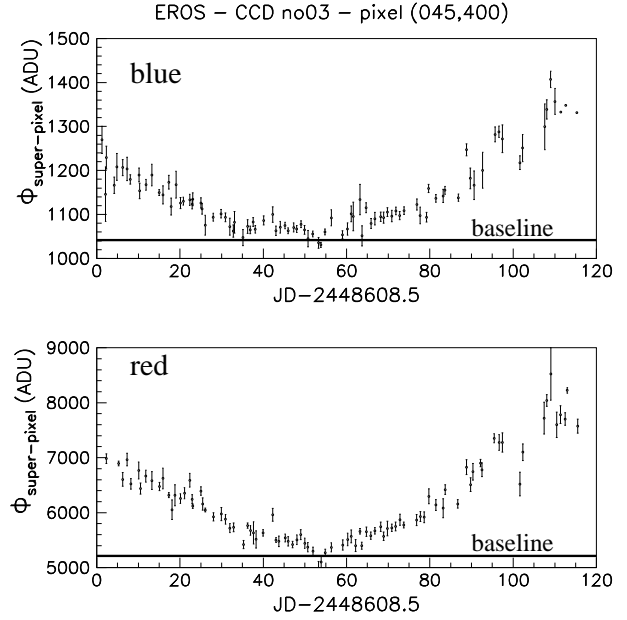}}
\caption{Super-pixel light curve with several variations rejected with
the cut on \protect $L_2$ ($L^B_1 = 1366.$ ; $L^R_1 = 2958.$ ; $L^B_2
= 565.$ ; $L^R_2 = 1655.$). The upper curve exhibits the light curve
in blue and the lower one in red. }
\label{fig:cl2b}
\end{figure}
At this stage, we have selected super-pixel light curves with at least
one significant variation. Now, it is important to check uniqueness.
We then require the second most significant fluctuation to have $L_2<
250$ in both colour. Fig.~\ref{fig:cl2b} shows an example of rejected
light curves with two variations.

\subsection{Correlation between the two colours}
\label{sect:corr} \label{section:correlation}
\begin{figure}
\resizebox{\hsize}{!}{\includegraphics{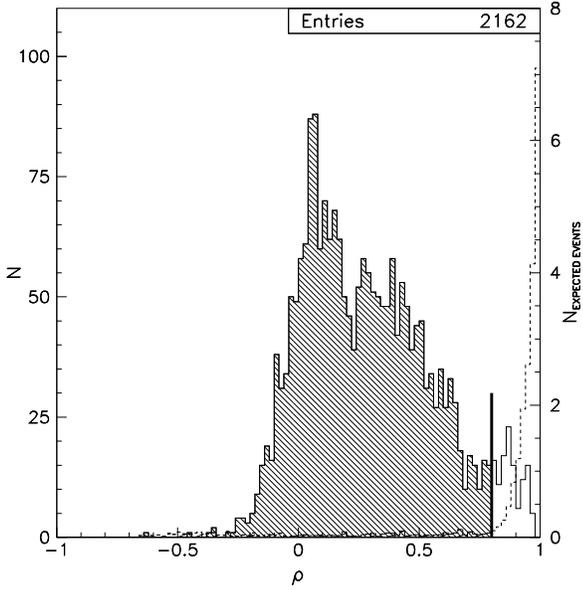}}
\caption{Histogram of the correlation factor computed between the two
colours. The hatched area corresponds to super-pixel light curves with
a correlation factor below the threshold \protect $\rho = 0.8$. The
histogram superimposed with a dashed line on this figure corresponds
to the simulations and scales with the tick marks and values given on
the right axis.}
\label{fig:cvrai}
\end{figure}
\begin{figure}
\resizebox{\hsize}{!}{\includegraphics{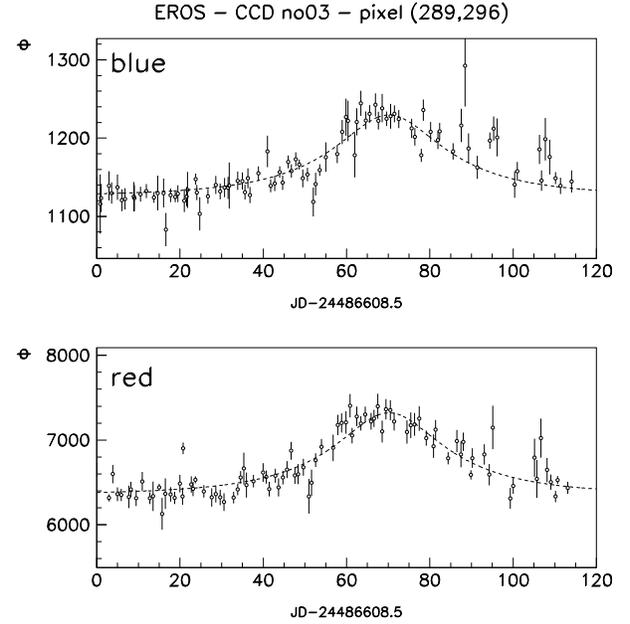}}
\caption{Super-pixel light curve with a shape compatible with a
microlensing event.  ($L^B_1 = 641. $ ; $L^R_1 = 744.$ ; $L^B_2 = 57.$
; $L^R_2 = 0.$ ; $\rho =0.84$)}
\label{fig:CL_3_289_296}
\end{figure}
\begin{figure}
\resizebox{\hsize}{!}{\includegraphics{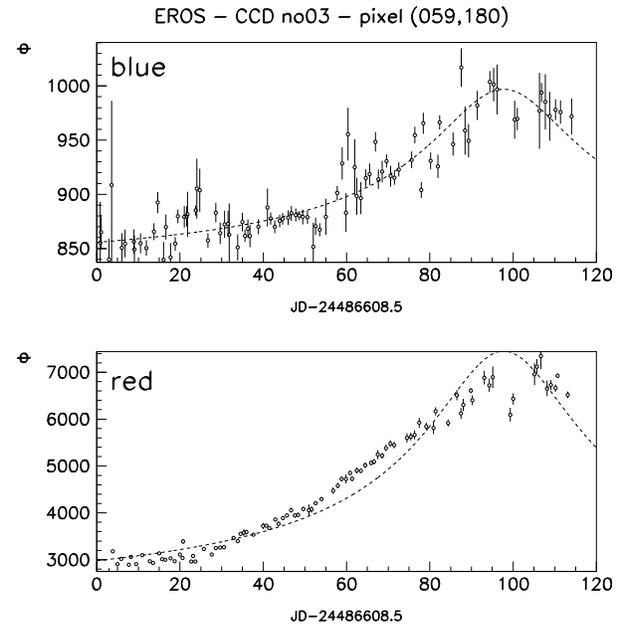}}
\caption{Super-pixel light curve with a shape incompatible with a
microlensing event ($L^B_1 = 1386. $ ; $L^R_1 = 16416.$ ; $L^B_2 =
80. $ ; $L^R_2 = 0. $ $\rho =0.93$)}
\label{fig:CL_3_059_180}
\end{figure}
\begin{figure}
\resizebox{\hsize}{!}{\includegraphics{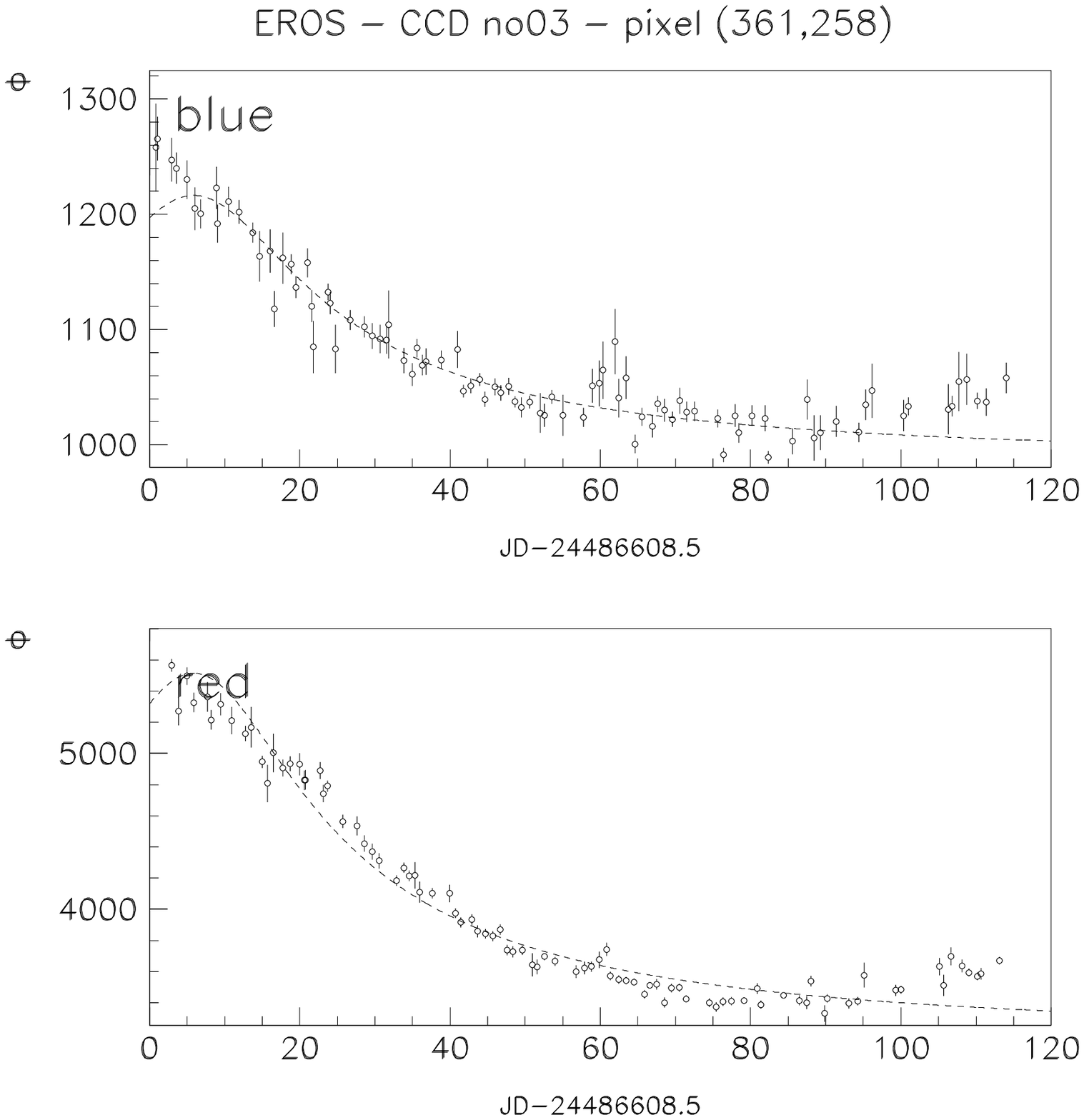}}
\caption{Super-pixel light curve for which it is inconclusive.  ($L^B_1 =
1791. $ ; $L^R_1 = 12927.$ ; $L^B_2 = 0.$ ; $L^R_2 = 176.$ $\rho
=0.96$)}
\label{fig:CL_3_361_258}
\end{figure}
In order to select the expected long time-scale variations, we choose
for the next requirement a reasonable correlation between the blue and
red light curves.

Given the constraints mentioned above, the correlation $\rho^p_{\rm
col}$ between the two colours:
\begin{equation}
\rho^p_{\rm col} = 
\frac{\sum_n {\left( \phi^p_n - {\langle \phi \rangle} \right) }
_{\rm blue} 
{\left( \phi^p_n - {\langle \phi \rangle} \right) }_{\rm red} }
{ \sqrt { { \sum_n {\left( \phi^p_n - {\langle \phi \rangle} \right)}
^2 }_{\rm blue} \sum_n {{ \left( \phi^p_n - {\langle \phi
\rangle}\right)}^2 }_{\rm red} }}
\label{eq:corr}
\end{equation}
achieves a good sensitivity to the achromaticity and to the dispersion
of the measurements.  As shown by the mock light curves in
Fig.~\ref{fig:cvrai} (dashed curve), a clear correlation between the
two colours is expected: most of the mock curves (85.5\%) lie above a
threshold $\rho = 0.8$.  Fig.~\ref{fig:cvrai} displays the
corresponding histogram for the {\em real light} curves selected so
far (full line). This distribution is quite different and exhibits a
peak close to $\rho = 0$.  With a threshold at $\rho^p_{col} > 0.8$,
94.4\% of the remaining light curves are removed: $120$ variations
remain.  Fig.  \ref{fig:CL_3_289_296} displays one of these light
curves: its shape is in quite good agreement with what we can expect
from a microlensing event although the period of observation is short
compared to the duration of the variation, and it is not possible to
test up to now the uniqueness of the variation.
Fig.~\ref{fig:CL_3_059_180} displays another light curve, one whose
shape is clearly incompatible with a standard microlensing
event. Fig.~\ref{fig:CL_3_361_258} shows a light curve for which it is
impossible to draw any conclusion based only on compatibility with the
microlensing shape: the period of observation is much shorter than the
time-scale of the variation.

When the criteria described above are applied in the simulations, we
expect $0.38 \times f$ events for $0.5 M_\odot$ Machos filling a
fraction $f$ of the halo. It is obvious then that the $120$ selected
light curves are in clear excess with respect to what is expected and
need to be further studied.

\section{Colour magnitude diagram}
\label{section:colmag}
\begin{figure}
\resizebox{\hsize}{!}{\includegraphics{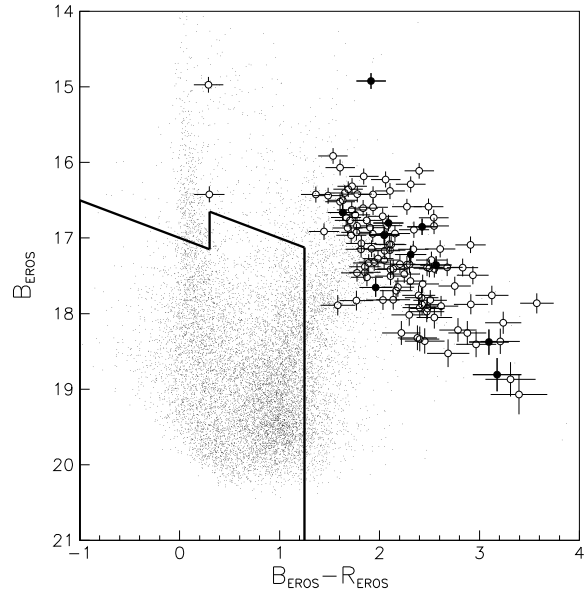}}
\caption{Colour magnitude diagram: the small dots corresponds to the
stars detected by the EROS-1 star monitoring analysis (Renault,
1996)\protect\nocite{Renault:1996}.  The circles corresponds to the
120 selected variations discussed in the text. Among them the filled
circles are Miras detected by Hughes (1989) in the I band (see text).}
\label{fig:colmag}
\end{figure}
Since the expected background for microlensing events is due to
variable stars, we have first to determine the position of the $120$
light curves with variations on the colour magnitude diagram (CMD),
before drawing any conclusion on the nature of these variations. The
problem is now to know how we can estimate the magnitude of the
underlying star responsible for the observed variation.

\subsection{Magnitude determination}
Using an image with a seeing close to the mean value ($3$ arc-second)
-- for which no significant seeing correction is required -- we
perform photometry on the surrounding stars using DAOPHOT which
returns the total background lying within $2$ arc-second of the center
of the super-pixel.  We then subtract this background from the
super-pixel flux measured when the star is at maximum luminosity, and
account for the seeing fraction of the star flux entering the
super-pixel. Hence, we deduce the magnitude of the star at the
maximum.  Due to the crowding of the \object{LMC Bar}, this aperture
photometry is the most efficient way to estimate a magnitude for the
stars responsible of the detected variations. More details about
magnitude determination will be further addressed in Melchior et
al. (1998b)\nocite{Melchior:1998c}.

 This estimate is mainly intended to study the position in the CMD of
the dominant source of flux of the varying pixel and in particular if
it lies in marginal locations of the colour-magnitude diagram,
characteristic of variable stars.  We estimate the uncertainties on
this magnitude determination as the square root of the sum of the
squares of the two following components.  The first is the error on
the super-pixel flux. The second one is estimated as 10\% of the
``star'' flux and is expected due to uncertainties in the star
position inside the inner pixel of the super-pixel.  In extreme cases
-- when the star flux at maximum is not the main contribution of the
super-pixel flux -- these errors can be underestimated.

\subsection{Discussion}
\begin{figure}
\resizebox{\hsize}{!}{\includegraphics{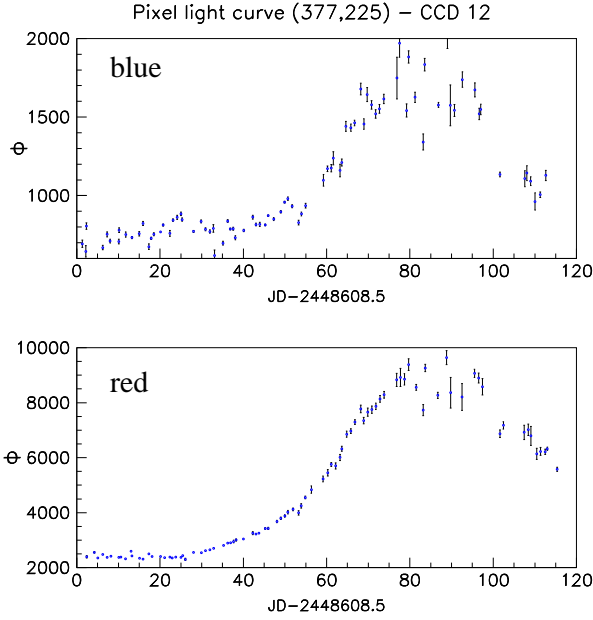}}
\caption{Super-pixel light curve with a variation detected by our
analysis which has been identify as a Mira. ($L^B_1 = 10188. $ ;
$L^R_1 = 47940.$ ; $L^B_2 = 112. $ ; $L^R_2 = 0. $ $\rho =0.96$)}
\label{fig:CL_12_377_225}
\end{figure}
\begin{figure*}
\resizebox{12cm}{!}{\includegraphics{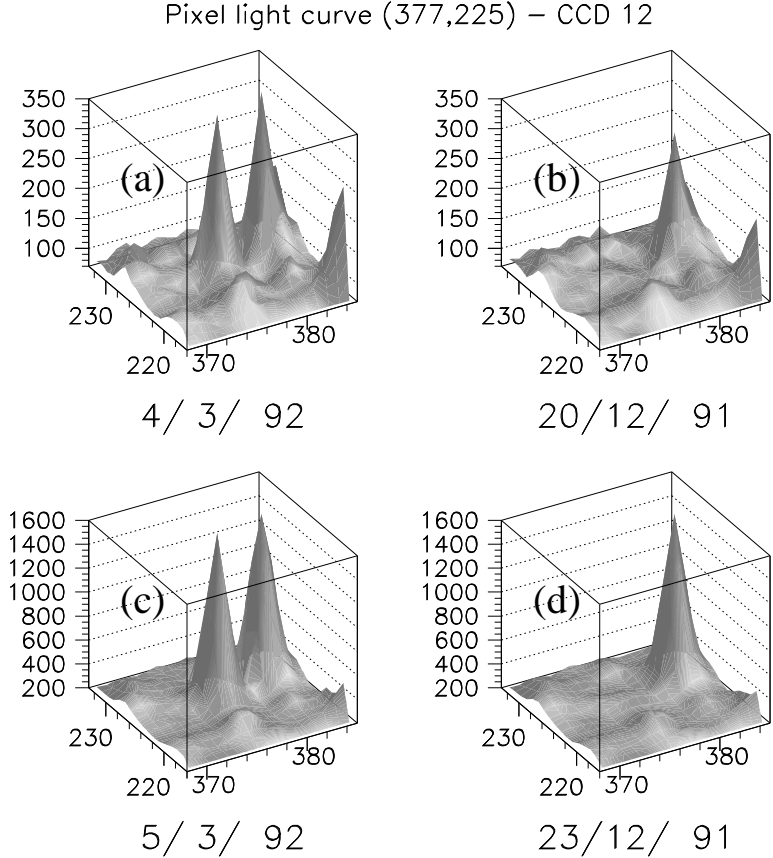}}
\hfill
\parbox[b]{55mm}{
\caption{Map of the surroundings of the Mira whose light curve is
displayed on Fig.~\protect \ref{fig:CL_12_377_225}: each figure
displays the flux as a function of position on an area of $20.6\arcsec
\times 20.6\arcsec$. Upper figures {\bf a} and {\bf b} correspond to
blue images, and lower ones {\bf c} and {\bf d} to red images. Figures
{\bf a} and {\bf c} on the left side exhibit the star at the maximum
of the variations. On the figures on the right side {\bf b} and {\bf
d}, this star has disappeared in the background. Note that such an
event would have escaped any star monitoring, and could not have been
accounted for as part of a blend.}
\label{fig:mirafield}}
\end{figure*}
\begin{table}
\caption[ ]{Miras: matches with a previous study by Hughes (1989). The
following characteristics are given: the position at 1950.0 epoch, the
CCD number, the pixel position, the mean $I$ magnitude, the period
in days
}
\label{tab:miras}
\begin{flushleft}
\begin{tabular}{cccccc}
\noalign{\smallskip}\hline\noalign{\smallskip}
$\alpha$ $\delta^a$ & CCD$^b$ & ix iy$^b$ & $I^a$ &
Period$^a$\\\noalign{\smallskip} \hline\noalign{\smallskip}
5.:16.:50.9 -69.:37.:52.0 &   0  &  295 220$^{~}$ & 13.87 & 170\\
5.:16.:57.0 -69.:19.: 9.0 &   8  &  ~93 545$^{~}$ & 14.36 & 183\\
5.:18.:19.7 -69.:41.:30.0 &   1  &  301 125$^{~}$ & 14.04 & 296\\
5.:20.:15.8 -69.:30.:59.0 &  10  &  252 173$^{~}$ & 14.96 & 293\\
5.:20.:35.5 -69.:43.:22.0 &   3  &  ~59 180$^c$ & 13.21 & 650\\
5.:21.: 8.7 -69.:37.:37.0 &   3  &  135 493$^d$ & 13.11 & 453\\
5.:21.:43.1 -69.:43.:28.0 &   3  &  346 240$^{~}$ & 13.94 & 453\\
5.:21.:47.6 -69.:43.:12.0 &   3  &  361 258$^{~}$ & 14.90 & 210\\ 
5.:23.:13.6 -69.:38.:45.0 &   4  &  258 556$^{~}$ & 13.77 & 255\\
5.:23.:52.9 -69.:34.:12.0 &  12  &  377 225$^e$ & 14.40 & 163\\
\noalign{\smallskip}
\hline
\end{tabular}
\\

$^a$Source: Hughes (1989)\nocite{Hughes:1989}
$^b$Source: AGAPEROS, this analysis
$^c$Cf. Fig.~\ref{fig:CL_3_059_180} 
$^d$Cf. Fig.~\ref{fig:CL_3_361_258}
$^e$Cf. Figs. \ref{fig:CL_12_377_225} and \ref{fig:mirafield}
\end{flushleft}
\end{table}
Fig.~\ref{fig:colmag} displays a CMD with the stars detected by the
EROS group (dots) and the underlying stars (circles) associated with
the selected super-pixel light curves. The variations kept by our
analysis are not representative of the bulk of the stars: all of them
but two lie in the red part of the CMD.

\paragraph*{The 118 red variable stars}\hspace{0.5cm}
The red detected variable stars are all located in the same area of
the colour magnitude diagram. We have even been able to check that
$10$ of them have already been recorded by Hughes
(1989)\nocite{Hughes:1989} as Miras with a study in the I band.  They
are displayed with filled circles on Fig \ref{fig:colmag} and their
known characteristics are displayed in Tab.~\ref{tab:miras}.  Fig.
\ref{fig:CL_12_377_225} exhibits the light curve of one of these
Miras, with a $163$ days period. Not surprisingly the variation is
significant but only sampled over $120$ days, and this is the shortest
period of the Miras listed in Tab.~\ref{tab:miras}. Two of these Miras
have been shown previously. One is presented in
Fig.~\ref{fig:CL_3_059_180}, it has a period of $650$ days. Another
one is exhibited in Fig \ref{fig:CL_3_361_258} with a period of $453$
days.  The catalogue of Hughes (1989)\nocite{Hughes:1989} contains
$41$ variable stars overlapping the field studied here and most of
them have been rejected at an earlier stage of this analysis.  The
thick lines show the area of the CMD excluded by the EROS group
corresponding to the regions where variable stars are expected. It is
then highly probable that the other red variable stars selected are
also Long Period Variables, as they lie in the same area of the
CMD. The fact that the majority of the red variable stars selected by
this analysis have not been identified previously demonstrates the
potential interest of the pixel method for the detection of Long
Period Variable (LPV) stars with respect to classical analysis
restricted to the study of resolved stars.  A comprehensive analysis
of these variable stars rejected as background of the microlensing
search will be presented elsewhere (Melchior et
al. 1998b)\nocite{Melchior:1998c}.

About 10 of these red variable stars lie below the crowding limit and
Fig.~\ref{fig:CL_12_377_225} displays the light curve of one of them:
the Mira already discussed above.  For these stars, we are not able to
detect variations around their minimum flux.  This is illustrated in
Fig.~\ref{fig:mirafield} which shows the field surrounding this star:
an unresolved star has exhibited a variation. Although this particular
example would have escaped the EROS-1 star monitoring applied on the
same data, it has already been identified in another wavelength (I
band) by Hughes (1989).

\paragraph*{The 2 blue variable stars}\hspace{0.5cm}
\begin{figure}
\resizebox{\hsize}{!}{\includegraphics{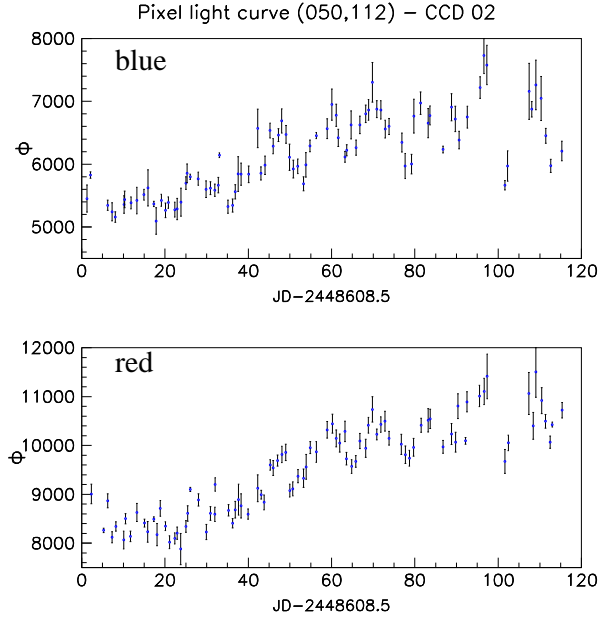}}
\caption{Super-pixel light curve with a variation affecting a blue
star detected by our analysis. It has been detected previously by
Beaulieu et al.\ (1996) and identified as a pre-main-sequence star
candidate labelled \object{ELHC2}. ($L^B_1 = 1747. $ ; $L^R_1 = 3850.$
; $L^B_2 = 0. $ ; $L^R_2 = 146. $ $\rho =0.89$)}
\label{fig:CL_02_050_112}
\end{figure}
The brightest blue variable star (with $B_E\simeq 15$) shown in
Fig.~\ref{fig:CL_02_050_112} belongs to the sample of
pre-main-sequence stars selected by Beaulieu et al.\
(1996)\nocite{Beaulieu:1996}. The short time scale variation which
superimposes on long-time scale 0.3 magnitude variation is real and
this feature excludes the simple microlensing interpretation anyway.

The other blue variable star is characterised by a small amplitude
($\Delta B_E < 0.3$) and a duration longer than $120$ days. These
features together with the position of this variable in the CMD
indicate that the variation is compatible with the new class of
variable stars called the blue bumpers identified by the MACHO group
(Cook et al. 1995, Alcock et
al. 1996)\nocite{Cook:1995b,Alcock:1996a}.

\subsection{Final cut on the colour magnitude diagram}
These considerations provide convincing evidences that the selected
variations are variable stars (Long Period Variables for most of
them).  We decide to apply the same cut as EROS on the CMD, displayed
with the thick lines shown in Fig.~\ref{fig:colmag}. These regions are
{\it filled with a negligible number of stars ($<$ 1.3\%)}, that are
moreover expected to be variable. The elimination of this area does
not significantly affect our sensitivity to microlensing events which
are expected to occur independently of the star's position in the
colour-magnitude diagram.

\section{Results}
\label{section:detect}
\begin{table}
\caption[ ]{Percentage of events kept for each step of the selection
procedure for different Macho's mass. The last row provides the
percentage of events kept with respect to the initial set of events.}
\label{tab:simu0}
\begin{flushleft}
\begin{tabular}{cccccc}
\noalign{\smallskip}
\hline
\noalign{\smallskip}
$M / M_\odot$ & $0.01$ & $0.05$ & $0.1$ & ${\mathbf 0.5}$ & $\mathbf
1$\\	  
\noalign{\smallskip} \hline\noalign{\smallskip}
$L_1 > 500$ in at& 32.0\%  & 42.1\%  & 
43.7\% & 48.9\% & 48.4\% \\
least one colour& & & & &\\\noalign{\smallskip}
 3 pts above 3$\sigma$ & 44.1\% & 58.6\%  & 
65.0\% & 71.2\% & 72.9\% \\
in both colours& & & & &  \\\noalign{\smallskip}
$L_2 < 250$ in & 96.4\% & 93.3\% & 92.9\% & 
89.7\% & 89.4\%\\   
both colours & & & & & \\\noalign{\smallskip}
$\rho > 0.8$ & 75.9\% & 76.0\% & 75.0\% & 
85.5\% & 88.7\% \\ \noalign{\smallskip} \hline\noalign{\smallskip} 
Total efficiency& 10.3\% & 17.4\%& 19.8\% & {\bf 26}.7\% & {\bf 27}.9\%
\\\noalign{\smallskip} \hline\noalign{\smallskip}  
\end{tabular}
\end{flushleft}
\end{table}
Although we detect no microlensing events with the selection procedure
described above, we would have detected them if there were
some. Whereas the detection of variable stars gives a first idea of
the sensitivity achieved by this analysis, the Monte-Carlo simulations
provide detection efficiencies.  Firstly, we present the detection
efficiencies achieved for this pixel analysis. Secondly, we compare
our sensitivity with those achieved by star monitoring analyses.

\subsection{Detection efficiencies for pixel monitoring} 
\begin{figure}
\resizebox{\hsize}{!}{\includegraphics{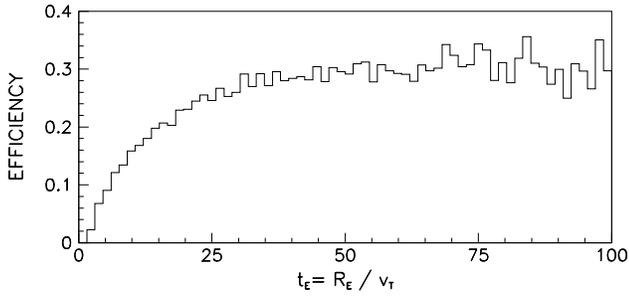}}
\caption{Detection efficiencies computed as a function of the Einstein
radius crossing time \protect${t_E}$.}
\label{fig:effte}
\end{figure}
In Sect. \ref{section:trigger}, we detail the effect of the selection
procedure on simulated light curves. For the clarity of the
discussion, we restricted then the comparison to events expected with
a halo full of $0.5 M_\odot$ Macho. We study here the sensitivity
achieved by our selection procedure on a larger-mass range.
Tab.~\ref{tab:simu0} shows the percentage of simulated events selected
by each step of our selection procedure for $ 0.01 M_\odot \le M \le 1
M_\odot$. It appears clearly that we have optimised the selection
procedure for this Macho mass range, and that efficiency is lost with
decreasing Macho mass.  The last cut on the correlation factor ($\rho
> 0.8$) is less favourable for $M\le 0.1M_\odot$: for variations
affecting only part of the period of observation, it is more optimal
to restrict the computation of this coefficient to the portions of the
light curve undergoing variations.
\begin{figure}
\resizebox{\hsize}{!}{\includegraphics{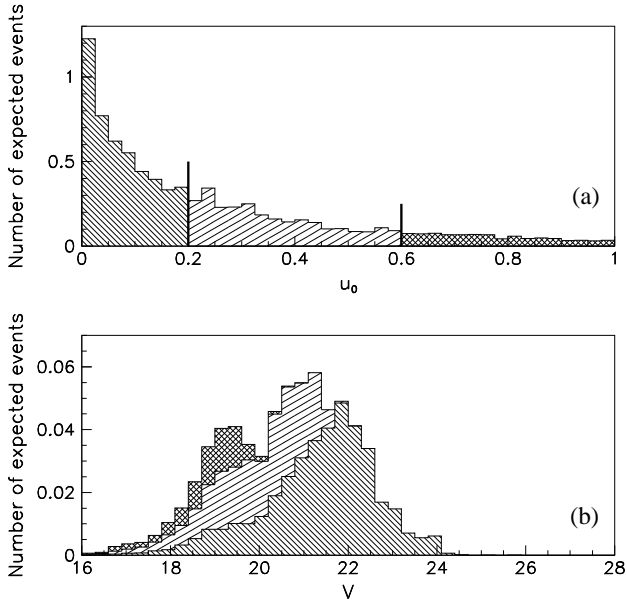}}
\caption{Selected simulated light curves. The distributions of the
impact parameter $u_0$ ({\bf a}) and of the $V$ magnitude ({\bf b}) of
the star at rest are exhibited, for 0.5 $M_\odot$
Machos. Normalisation corresponds to a full halo.}
\label{fig:para1}
\end{figure}
\begin{figure}
\resizebox{\hsize}{!}{\includegraphics{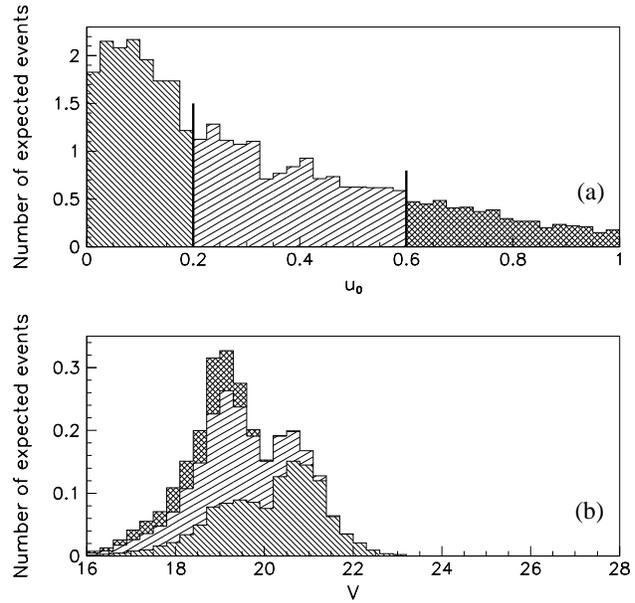}}
\caption{Selected simulated light curves. The distributions of the
impact parameter $u_0$ ({\bf a}) and of the $V$ magnitude ({\bf b}) of
the star at rest are exhibited, for 0.01 $M_\odot$
Machos. Normalisation corresponds to a full halo.}
\label{fig:para101}
\end{figure}
These efficiencies can be studied as a function of the duration
${t_E}$ as shown in Fig.~\ref{fig:effte}. Due to the temporal
sampling, the selection procedure is less efficient for short duration
events. The efficiency remains constant for long-duration events, as
we do not require a stable baseline.

Fig.~\ref{fig:para1} shows the distribution of impact parameters and V
magnitude for simulated events due to 0.5 $M_\odot$ Machos satisfying
all our requirements. It is clear that with respect to
Fig.~\ref{fig:para0}, the selection procedure eliminates microlensing
events affecting very dim stars ($V>24$), and that the main
contribution is expected due to events affecting dim stars with a
small impact parameter. Fig.~\ref{fig:para101} gives the same
histograms but for events due to $0.01 M_\odot$ Machos. Events
affecting dim stars are much more difficult to detect if they are
short. Hence, detected events affect on average brighter stars and the
impact parameter distribution appears flatter.
\begin{figure}
\resizebox{\hsize}{!}{\includegraphics{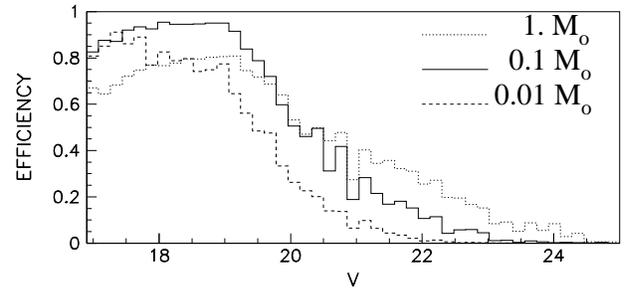}}
\caption{Detection efficiencies computed as a function of the
magnitude of the un-magnified star.}
\label{fig:effmv}
\end{figure}

Fig.~\ref{fig:effmv} exhibits the detection efficiencies achieved as a
function of the V magnitude of the un-magnified star and illustrates
the previous point. The longer the events, the dimmer the stars they
can affect. It is also to be noted that events affecting bright stars
can be missed when the event duration is longer than the period of
observation.

\subsection{Comparison with star monitoring}
\label{sect:compa}
\begin{table}
\caption[ ]{Number of expected microlensing events for this pixel
analysis and the EROS star monitoring analysis.  The 2nd row gives the
number of events \protect $N^{\rm AGAPEROS}_{\rm evt}$ expected given
these efficiencies and a halo mass fraction \protect $f$, that can be
compared with the number of events $N^{\rm EROS}_{\rm evt}/f$ expected
on the same data (Renault 1996) shown in the 3rd row. The last row
displays the ratio $N^{\rm AGAPEROS}_{\rm evt} / N^{\rm EROS}_{\rm
evt}$.}
\label{tab:simu1}
\begin{flushleft}
\begin{tabular}{cccccc}
\hline \noalign{\smallskip} 
$M / M_\odot$ & $0.01$ & $0.05 $ & $ 0.1$ & $ 0.5$ & $ 1 $\\ 
\noalign{\smallskip} \hline\noalign{\smallskip}
$N^{\rm AGAPEROS}_{\rm evt}/f$ &  $ 1.02 $ & $ 0.77 $ & $ 0.64 $ & $
0.38 $ & $ 0.28 $ \\
$N^{\rm EROS}_{\rm evt}/f$ &  $0.14$ & $0.055$ & $0.045$ & $0.018$ &
$0.014$\\ 
Ratio &   6 & 11 & 13 & 19 & 19
\\ \noalign{\smallskip}\hline\noalign{\smallskip} 
 \end{tabular}
\end{flushleft}
\end{table}
Tab.~\ref{tab:simu1} gives the number of microlensing events expected
with our AGAPEROS pixel analysis as well as the number of microlensing
events expected with the EROS star monitoring analysis. For
long-duration events due to $M \ge 0.5 M_\odot$ Machos in the mass
range where our analysis has been optimized, the number of
microlensing events that our pixel analysis could detect is enhanced
by a factor larger than 15. This gain is due to the fact that the EROS
star monitoring analysis accounts for stars down to magnitude 19.5 but
is far from complete down to the limiting magnitude, given the
crowding and seeing conditions. In the LMC bar fields, stellar
photometry has a bad detection efficiency for dim stars, whereas our
approach does not require to resolve the stars to detect their
variation.

Similar considerations can be made with respect to the MACHO star
monitoring analysis (Alcock et al. 1996\nocite{Alcock:1996a}) applied
on a field 60 times larger than the 0.25 deg$^2$ field analysed
here. The MACHO exposure was $E_{MACHO}=9.7 \times 10^6$
star-yr. corresponding to an experiment duration of 409 days. In first
approximation, the equivalent MACHO exposure for 0.25 deg$^2$ and 120
days would be $E^\prime_{MACHO} = 4.7 \times 10^4 $ star-yr. As the
detection efficiencies present similar features as those discussed
here, this exposure can be compared with the AGAPEROS one
$E_{AGAPEROS}= N^{\rm AGAPEROS}_{\rm stars} \times T_{\rm obs}(yr) =
7.2 \times 10^5$ star-yr, which is about 15 times larger.

\section{Perspectives for the Magellanic Clouds}
\label{section:perspec}
 We discuss here the perspectives than can be expected with the whole
EROS-1 CCD data set (Sect. \ref{sect:eros1}), then with other data
characteristics (Sect. \ref{sect:other}), and give general
considerations about further applications of this technique
(Sect. \ref{sect:pixmeth}).

\subsection{The whole EROS-1 CCD data set}
\label{sect:eros1}
The limitation of this analysis due to the short-time span over which
the 91-92 data stretched could be overcome by analysing the whole
EROS-1 CCD data set. We discuss here the perspectives offered by such
an analysis.  We use the same simulations, but with the
characteristics of the $6871$ images available in blue and $7011$
images in red, namely the time, absorption factor and background
flux. Tab.~\ref{tab:9194} shows the number of events expected
assuming that the same selection procedure\footnote{In practice, the
thresholds may have to be adjusted to account for possible unexpected
sources of noise, hence changing the sensitivity.} can be used for the
whole EROS-1 CCD data set.
\begin{table}
\caption[ ]{Simulations for the EROS 91-94 data: the number of events
$N^{\rm AGAPEROS}_{\rm evt} / f $ expected with a pixel analysis
applying the same selection procedure as described above. The last row
gives for comparison the number of expected events $N^{\rm EROS}_{\rm
evts}$ with the star monitoring.}
\label{tab:9194}
\begin{flushleft}
\begin{tabular}{ccc}
\hline
\noalign{\smallskip}
$M$ &  $0.1 M_\odot$ & $0.5 M_\odot$ \\
\noalign{\smallskip} \hline\noalign{\smallskip}
$N^{\rm AGAPEROS}_{\rm evts}/f$ (pixel monitoring) & $ 5.0 $ & $3.0$ \\
$N^{\rm EROS}_{\rm evts}/f$ (star monitoring) & $ 0.42 $ & $0.2$ \\
\noalign{\smallskip} \hline
\end{tabular}
\end{flushleft}
\end{table}
Hence, we can reasonably expect $\simeq 4 \times f$ events for $0.1 -
0.5 M_\odot$ Machos. This sensitivity is equivalent to the one
typically achieved using a star monitoring analysis performed on a
much larger field (EROS-1 plates: 25 deg$^2$ - around 10$\times f$
expected events for 0.1 $M_\odot$ Machos).

\subsection{Other sources of data}
\label{sect:other}
The pixel analysis described in this paper has been applied on data
which were rather peculiar among the microlensing databases. An
average of 10 exposures per night was available: this is the reason
why we succeed to achieve a relative stability between 1 and 2\% on
super-pixels (see Paper~I). For comparison, the typical level of
photon noise obtained on the background of the EROS-2 images on the
\object{LMC Bar} is about 0.3\% (for super-pixels). Then the stability
that could be achieved on corresponding super-pixel light curves could
be typically twice the photon noise, namely about 0.6\%.  Similar
stability could be expected with MACHO images (D. Bennett, private
communication).

One must also keep in mind that this analysis has been applied to a
field in the Bar of the \object{LMC}, which is a very crowded
field. The gain of pixel monitoring would be substantially smaller
when used in less dense regions.

\subsection{Pixel Method}
\label{sect:pixmeth}
The principal problem with a pixel analysis towards the \object{LMC}
is that, by its very nature, it has difficulty measuring the flux of
the un-magnified star and the maximum magnification. One significant
consequence of this problem is that it produces degeneracies that
affect the determination of the duration ${t_E}$ of the events.

However, the same problem arises with star monitoring which is
affected in a major way by blending: that is, a parameterisation
similar to the one presented in Eq.~\ref{eq:phi} must also be
considered (see Wozniak \& Paczy\'nski\ 1997, Alcock et al.\
1996\nocite{Wozniak:1997, Alcock:1996a}) to account for the
magnification of underlying stars.

The MACHO and EROS groups (Alcock et al.\ 1997a, Pratt 1997, Renault
et al. 1998, Palanque-Delabrouille et al.\
1998)\nocite{Alcock:1997a,Pratt:1997,Renault:1998,Palanque-Delabrouille:1998}
has corrected their events (detected by star monitoring) for blending
effects with a statistical correction.  Moreover, one could in
principle overcome this difficulty with a high resolution image
achieving a good signal to noise ratio down to $V=24$ (see Han
1997)\nocite{Han:1997}. Ardeberg et al. (1997)\nocite{Ardeberg:1997}
have measured the flux of the stars in the \object{LMC Bar} on HST
images down to magnitude 24 (Str\"omgren photometry) and claimed to be
completed down to magnitude 22.  Such a performance should solve the
problem of measuring the flux of the un-magnified star for most of the
events (see Fig.~\ref{fig:para1}). However, the identification of the
star that has varied is one possible practical problem which must be
studied further, but could probably be overcome.  For the dimmest
stars, unambiguous determination of the stellar flux will most
probably require an HST measurement of the star flux during the event.
Such a requirement seems reasonable for an ambitious pixel experiment
towards the \object{LMC}: HST measurements are already being performed
to correct for blending in events detected by star monitoring toward
the bulge.

\section{Conclusions}
While the EROS-1 CCD data set has already excluded the small-mass
Macho range, we have shown that it is possible with a pixel analysis
of the same data set to probe the mass range ($0.01 - 1. M_\odot$)
where all the known events have been detected.  We thus demonstrate
for the first time the efficacy of the pixel method for \object{LMC}
images.  We have shown with the computation of detection efficiencies
that the gain in detectable microlensing events with pixel monitoring
is significant: {\em pixel monitoring can detect $15$ times more
microlensing events than star monitoring} for very crowded LMC bar
fields.

With simple selection criteria we have been able to show that no
microlensing events are present in this data set. The criteria imposed
mainly rely on the uniqueness of the variation, on the achromaticity
and on the fact that the microlensing phenomenon -- a geometrical
effect -- should affect all kind of stars independently of their
type. This demonstrates that the noise which affects super-pixel light
curves can be reduced to a level that is adequate to conduct the
analysis.  Our analysis, optimised for the detection of long duration
events, has only detected variable stars, and we have shown our
ability to reject such variable stars with the CMD. We have then
demonstrated the efficacy of this approach for the detection of
variable stars, which remain the main background for microlensing.  A
comprehensive analysis of the variable stars rejected by this analysis
is in progress (Melchior et al. 1998b)\nocite{Melchior:1998c}.

Such an analysis can detect microlensing magnification of unresolved
stars up to $V=24$ and is thus complementary to star
monitoring. On-line pixel analysis and follow-up will probably be
necessary to better discriminate possible variable stars and to
achieve a higher quality photometry.

The short period of observation analysed here -- $120$ days to be
compared to $3$ years for EROS-1 plates and $2$ years for MACHO -- and
the relatively small field -- $100$ times smaller than the one
analysed by the EROS group (plates) (Ansari et
al. 1996)\nocite{Ansari:1996} and $60$ times smaller than the field
analysed by the MACHO group (Alcock et al.\
1997a)\nocite{Alcock:1997a} -- explain the relatively small number of
events expected. For the first time, we are able to provide detection
efficiencies for a pixel analysis. These results allow us to estimate
that a pixel analysis of the complete existing EROS-1 CCD data base
could detect about $\simeq 4. \times f$ microlensing events in the
mass range of interest ($0.1 - 0.5 M_{\odot}$)... with a field of only
$0.39\,\rm deg^2$!

\acknowledgements{We are particularly grateful to C. Lamy for her
useful help on data handling during this work. We thank S. Hughes for
its interest for this work and his useful contribution for
Fig.~\ref{fig:colmag}. We thank the MACHO collaboration and in
particular D. Bennett and M. Pratt for kindly providing their
detection efficiencies.  During this work, A.L. Melchior has been
supported by a grant from Singer-Polignac Foundation, a grant from
British Council and PPARC and a European contract ERBFMBICT972375 at
QMW.}

\end{document}